%% file: main.tex
\documentclass[sigconf]{acmart}

\AtBeginDocument{%
  \providecommand\BibTeX{{%
    \normalfont B\kern-0.5em{\scshape i\kern-0.25em b}\kern-0.8em\TeX}}}

\copyrightyear{2022}
\acmYear{2022}
\setcopyright{acmcopyright}\acmConference[CIKM '22]{Proceedings of the 31st ACM International Conference on Information and Knowledge Management}{October 17--21, 2022}{Atlanta, GA, USA}
\acmBooktitle{Proceedings of the 31st ACM International Conference on Information and Knowledge Management (CIKM '22), October 17--21, 2022, Atlanta, GA, USA}
\acmPrice{15.00}
\acmDOI{10.1145/3511808.3557445}
\acmISBN{978-1-4503-9236-5/22/10}

\usepackage{amsmath}
\usepackage{acronym}
\usepackage{multirow}
\usepackage{enumitem}
\usepackage{tablefootnote}
\usepackage{bbm}
\usepackage{multirow}

\acrodef{PTM}{pre-trained model}
\acrodef{NRM}{neural ranking model}

\makeatletter
\g@addto@macro\normalsize{%
  \abovedisplayskip 2pt plus1pt 
  \belowdisplayskip 2pt plus1pt
  \abovedisplayshortskip  2pt plus1pt%
  \belowdisplayshortskip  2pt plus1pt
}
\makeatother

\begin{document}

\title{Scattered or Connected? An Optimized Parameter-efficient Tuning Approach for Information Retrieval}

\author{Xinyu Ma}
\affiliation{
 \institution{CAS Key Lab of Network Data Science and Technology, ICT, CAS}
 \institution{University of Chinese Academy of Sciences}
 \city{Beijing}
 \country{China}
}
\email{maxinyu17g@ict.ac.cn}

\author{Jiafeng Guo}
\authornote{Jiafeng Guo is the corresponding author.}
\affiliation{
 \institution{CAS Key Lab of Network Data Science and Technology, ICT, CAS}
 \institution{University of Chinese Academy of Sciences}
 \city{Beijing}
 \country{China}
}
\email{guojiafeng@ict.ac.cn}
 
\author{Ruqing Zhang}
\affiliation{
 \institution{CAS Key Lab of Network Data Science and Technology, ICT, CAS}
 \institution{University of Chinese Academy of Sciences}
 \city{Beijing}
 \country{China}
}
\email{zhangruqing@ict.ac.cn}

\author{Yixing Fan}
\affiliation{
 \institution{CAS Key Lab of Network Data Science and Technology, ICT, CAS}
 \institution{University of Chinese Academy of Sciences}
 \city{Beijing}
 \country{China}
}
\email{fanyixing@ict.ac.cn}
 
\author{Xueqi Cheng}
\affiliation{
 \institution{CAS Key Lab of Network Data Science and Technology, ICT, CAS}
 \institution{University of Chinese Academy of Sciences}
 \city{Beijing}
 \country{China}
}
\email{cxq@ict.ac.cn}

\renewcommand{\shortauthors}{Xinyu and Ruqing, et al.}

\begin{abstract}

Pre-training and fine-tuning have achieved significant advances in the information retrieval (IR). 
A typical approach is to fine-tune all the parameters of large-scale pre-trained models (PTMs) on downstream tasks. 
As the model size and the number of tasks increase greatly, such approach becomes less feasible and prohibitively expensive. 
Recently, a variety of parameter-efficient tuning methods have been proposed in natural language processing (NLP) that only fine-tune a small number of parameters while still attaining strong performance.
Yet there has been little effort to explore parameter-efficient tuning for IR. 

In this work, we first conduct a comprehensive study of existing parameter-efficient tuning methods at both the retrieval and re-ranking stages. 
Unlike the promising results in NLP, we find that these methods cannot achieve comparable performance to full fine-tuning at both stages when updating less than 1\% of the original model parameters. 
More importantly, we find that the existing methods are just parameter-efficient, but not learning-efficient as they suffer from unstable training and slow convergence.  
To analyze the underlying reason, we conduct a theoretical analysis and show that the separation of the inserted trainable modules makes the optimization difficult.
To alleviate this issue, we propose to inject additional modules alongside the \acp{PTM} to make the original scattered modules connected.
In this way, all the trainable modules can form a pathway to smooth the loss surface and thus help stabilize the training process.
Experiments at both retrieval and re-ranking stages show that our method outperforms existing parameter-efficient methods significantly, and achieves comparable or even better performance over full fine-tuning.

\end{abstract}



\maketitle

\section{Introduction}

``Pre-training and fine-tuning'' has became the prevalent paradigm in the natural language processing (NLP) \cite{Qiu2020ptm4nlp,bommasani2021ftm}.  
The success of Transformer-based pre-trained models (\acp{PTM}) in the NLP has also attracted attention in the information retrieval (IR)  community~\cite{Lin2021bert-survey,yixing-ptm4ir}. 
Many researchers have applied the popular \acp{PTM}, e.g., BERT \cite{Devlin2019BERT} and RoBERTa \cite{liu2019roberta}, into the multi-stage search pipeline \cite{ance,nogueira2019bert-pas-ranking}, including the first-stage retrieval and the re-ranking stage. 
The first-stage retrieval aims to return a subset of candidate documents efficiently, and the re-ranking stage attempts to re-rank those candidates accurately.  
Studies have shown that leveraging the existing \acp{PTM} can benefit both the retrieval and re-ranking stages significantly~\cite{nogueira2019bert-pas-ranking,Karpukhin2020dpr,Zhan2021adore,Khattab2020colbert}.

The mainstream approach to adapt large-scale \acp{PTM} to the downstream tasks is via full fine-tuning, which updates all the parameters of the \acp{PTM}. 
Though effective, this fine-tuning approach has drawbacks on its parameter efficiency. 
Firstly, every downstream task needs a separate copy of fine-tuned model parameters, containing as many parameters as in the original \acp{PTM}. 
This is prohibitively expensive when serving models that perform a wide range of tasks. 
Secondly, larger models are usually trained every few months with the ever increasing size ranging from millions~\cite{Devlin2019BERT} to hundreds of billions \cite{GPT3} or even trillions of trainable parameters~\cite{switch-transformer}. 
As the model size and the number of tasks grow, re-training all model parameters becomes less feasible and raises critical deployment challenges.

To alleviate this issue, a surge of development of parameter-efficient tuning methods have been proposed in NLP, which update only a small number of extra parameters while keeping the original \acp{PTM} parameters frozen  \cite{Schick2021prompt,houlsby2019adapter,prefix-tuning,zaken2021bitfit,he2021mam-adapter,lora}. 
The representative methods include addition-based such as Adapter~\cite{houlsby2019adapter} and prefix-tuning~\cite{prefix-tuning}, specification-based such as Bitfit \cite{zaken2021bitfit}, and low-rank adaption like LoRA~\cite{lora}. 
Most of these methods are injected to the \acp{PTM} in an inside manner where the extra tunable modules are scattered in the sub-layers of the Transformer.  
In essence, the inside modules have a big impact to the final output due to their interaction with the original \acp{PTM}. 
These methods have been reported to achieve comparable performance over full fine-tuning on NLP tasks, with only updating less than 1\% of the original parameters. 

Yet there has been little effort to adopt parameter-efficient tuning to the IR scenario. 
The most related work in this direction focused on the re-ranking stage \cite{Jung2022sime-siamese}, where prefix-tuning and LoRA are delicately leveraged. 
Their experimental results demonstrate that these two methods generally perform on par with or even outperform the full fine-tuning by tuning less than 1\% of the original model parameters. 
However, the retrieval stage remains less well studied.
Besides, the proposed mechanism is designed for the bi-encoder and unsuitable for the cross-encoder, resulting in the limitation of its flexibility.
In addition, their experimental results are on small test sets, which may not be representative enough.

In this work, we first conduct a comprehensive study of several representative parameter-efficient tuning methods for both the retrieval and re-ranking stage.   
The first research question is: can existing methods perform as well in IR as in NLP?
The results show that: 
(1) These methods lag behind full fine-tuning on both stages by tuning less than 1\% of orignal parameters, which is different from the findings observed from the small IR datasets \cite{Jung2022sime-siamese}. 
(2) Existing parameter-efficient tuning methods suffer from unstable training and slow convergence. 
That is, these methods are just parameter-efficient, but not learning-efficient.

This phenomenon raises the second research question: why the standard setup of parameter-efficient tuning methods falls short in IR? 
To analyze the underlying reason, we conduct a theoretical analysis and find the potential reason is that the separation of the inserted trainable modules results in a discrepancy between the ideal optimization direction and the actual update direction.
Specifically, the computation of the optimization direction depends on the parameters of the whole model (including the \acp{PTM} and injected modules), while the actual gradient update only performs on the injected modules. 
Such discrepancy makes the optimization difficult, which may hurt the performance.

The above analysis leads to the third research question: can we design a parameter-efficient tuning approach to stabilize the training process?  
Inspired by the skip connection \cite{he2016resnet} in deep learning, we propose to insert extra modules in an aside manner beyond the inside manner. 
The key idea is that extra modules injected alongside the \acp{PTM} could make the original scattered modules connected.
In this way, all the trainable modules can form a pathway to smooth the loss surface and thus help stabilize the training process. 
In this work, we carefully design three insertion ways of the aside module. 
By combing the inside and aside modules, our method can well inherit their advantages, i.e., smoothed loss of the aside modules and big impact of the inside modules.  
Note that our method can combine most of the parameter-efficient methods and is able to serve both the retrieval and re-ranking stage, and both cross-encoder and bi-encoder.   
Experiments at both retrieval and re-ranking stages show that our method is significantly better than existing parameter-efficient tuning methods. 
With tuning less than 1\% of the original parameters, our method can achieve comparable performance over full fine-tuning.
With tuning 6.7\% of the original parameters, our method is able to outperform the full fine-tuning on most tasks.

\input{preliminary}

\input{method}

\input{exp-settings}

\input{exp-results}

\input{related}

\section{Conclusion}
In this paper, we conduct comprehensive empirical studies of parameter-efficient tuning methods in IR scenarios, at both the retrieval stage and the re-ranking stage.
On four standard large-scale benchmarks, we find that these methods are unable to outperform or even achieve a comparable performance over full fine-tuning with tuning less than 1\% of original model parameters.
Through mathematical analysis, we certify the reason is that the separation of the trainable parameters results in a discrepancy between the ideal optimization direction and the actual update direction.
We thus introduce the aside module to help to stabilize the optimization process.
Experiments show that our method is significantly better than existing methods and could outperform the full fine-tuning on most tasks by tuning 6.7\% of original model parameters.
In future work, we would study their ability of domain adaptation in IR.

\begin{acks}
This work was funded by the National Natural Science Foundation of China (NSFC) under Grants No. 62006218 and 61902381, the Youth Innovation Promotion Association CAS under Grants No. 20144310, and 2021100, the Young Elite Scientist Sponsorship
Program by CAST under Grants No. YESS20200121, and the Lenovo-CAS Joint Lab Youth Scientist Project.
\end{acks}

\bibliographystyle{ACM-Reference-Format}
\balance
\bibliography{main}


\end{document}

%% file: preliminary.tex
\section{Preliminary}

In this section,  we give a brief description of the ranking problem in IR, the Transformer architecture, as well as several representative parameter-efficient tuning methods. 

\subsection{Problem Statement}
To balance the search efficiency and effectiveness, modern search systems typically employ a multi-stage ranking pipeline in practice, including the first-stage retrieval stage and the re-ranking stage~\cite{yixing-ptm4ir}.

\subsubsection{Dense Retrieval}
For the retrieval stage, the model needs to recall a small set of documents from a large-scale corpus efficiently. 
Dense retrieval models usually employ a representation-based architecture (i.e., bi-encoder) to encode queries and documents into low-dimensional representations independently~\cite{Karpukhin2020dpr,Ma2022COSTA}. 
Simple similarity functions like dot-product are adopted to compute the relevance score with the dense representations.

Without the loss of generality, the retrieval function with the representation-based architecture can be formulated as follows:
\begin{equation}
\label{equ-dense-retrieval}
rel(q,d) = f(\phi_{_{PTM}}(q), \varphi_{_{PTM}}(d)),
\end{equation}
where $\phi_{_{PTM}}$ and $\varphi_{_{PTM}}$ are query and document encoders, and $f$ is the similarity function.

\subsubsection{Re-ranking}

At the re-ranking stage, the interaction-focused model is widely adopted to produce more accurate ranking list \cite{MacAvaney2019cedr, Ma2021prop,B-Prop}. 
The relevance score is usually computed by a feed-forward neural network at the top of \acp{PTM} where queries and documents are concatenated together as the input to the model.

Without loss of generality, the re-ranking function with the interaction-based architecture could be abstracted as:
\begin{equation}
\label{equ-reranking}
rel(q,d) = f(\eta_{_{PTM}}(q,d))
\end{equation}
where $\eta_{_{PTM}}$ is the interaction function based on $\acp{PTM}$, and $f$ is the scoring function based on the interaction features.
Even though the representation-based models can also be applied to the re-ranking stage, studies have shown that they are less effective than the interaction-based models~\cite{qiao2019bert,sean-prettr}.

\begin{figure}[t]
    \centering
    \includegraphics[scale=0.35]{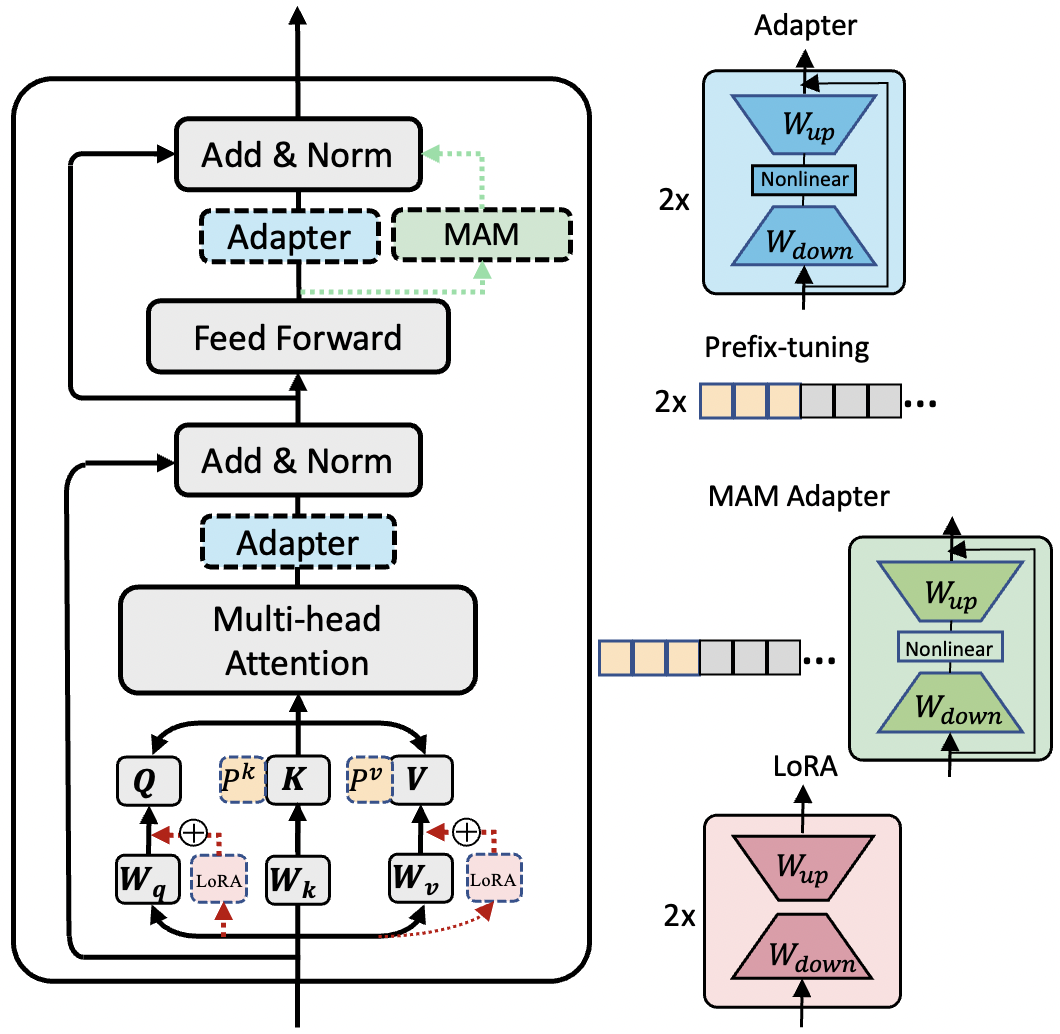}
    \caption{Illustration of a Transformer layer and several representative parameter-efficient tuning methods. 
    Note that MAM Adapter uses a parallel adapter on FFN sub-layer and prefix-tuning on self-attention sub-layer.}
    \label{fig:existing-methods}
\end{figure}

\subsection{Transformer}\label{sec:transformer}
Transformer is the dominant model architecture for \acp{PTM}. 
Specifically, a Transformer layer~\cite{vaswani2017attention} contains a self-attention sub-layer, a feed-forward neural network sub-layer, and residual connection followed by layer normalization.

\subsubsection{Self-Attention}
The input hidden states are firstly transformed to three vectors, i.e., queries, keys, and values, $m$ times independently where $m$ is the number of heads.
Then a dot-product function is applied on queries and keys to compute attention weights for each head, and then a weighted sum operation is performed on the values.
Given the hidden state $h\in\mathbb{R}^{n \times d}$, the $i$-th attention is computed as:
\begin{equation}
	\operatorname{Attention}_i(\textbf{h})=\sum_m\operatorname{softmax}(\frac{W_i^q\textbf{h}\cdot W_i^k\textbf{h}}{\sqrt{d/m}})W_i^v\textbf{h},
\end{equation}
where $W_i^q, W_i^k, W_i^v \in \mathbb{R}^{d/m \times m} $ are the learned transformation matrices for queries, keys and values. 

Finally, the output of the multi-head attention is computed as a concatenation of the output vectors of all the heads
\begin{equation}\label{eq:mh}
	\operatorname{MH}(\textbf{h})=\operatorname{Concat}(\operatorname{Attention}_1(\textbf{h}), \dots, \operatorname{Attention}_n(\textbf{h}))W^o,
\end{equation}
where $W^o \in \mathbb{R}^{d\times d}$ is the projection matrix. 

\subsubsection{Feed-forward Neural Network}
The feed-forward network is a position-wise fully connected feed-forward network (FFN), which is applied to each position separately and identically,
\begin{equation}\label{eq:ffn}
   FFN(\textbf{h})=ReLU(\textbf{h}W_1+b_1)W_2+b_2
\end{equation}
where $W_1 \in \mathbb{R}^{d\times 4d}$, $W_2 \in \mathbb{R}^{4d\times d}$, $b_1,$ and $b_2$ are learned bias terms.

Each of the two sub-layers, i.e., the self-attention sub-layer and the FFN sub-layer, employ a residual connection followed by layer normalization (RCLN) to compute the final output
\begin{equation}\label{eq:rcln}
	RCLN(h) = \operatorname{LayerNorm}(\operatorname{SubLayer}(\textbf{h}) +\textbf{\textbf{h}}),
\end{equation}
where $\operatorname{LayerNorm}(\cdot)$ is layer normalization and $\text{SubLayer}$ representations Eq.~(\ref{eq:mh}) and Eq.~(\ref{eq:ffn}).

\subsection{Parameter-efficient Tuning Methods}\label{sec:pet}

We introduce five representative parameter-efficient tuning methods as illustrated in Figure~\ref{fig:existing-methods}.
These methods can be categorized to three groups, i.e., Addition-based, Specification-based and Low-rank adaption.

\subsubsection{Addition-based}
Addition-based methods introduce extra parameters by inserting small neural modules such as Adapter~\cite{houlsby2019adapter}, or trainable tokens such as prefix-tuning~\cite{prefix-tuning}.
Only these additional parameters are tuned while the original parameters of \acp{PTM} are kept frozen.
Besides adapter and prefix-tuning, we also consider the recently proposed Mix-And-Match Adapter (MAM Adapter)~\cite{he2021mam-adapter}.

\begin{itemize}[leftmargin=*]
    \item Prefix-tuning extends the prompt-tuning~\cite{Lester2021prompt-tuning} by prepending $m$ trainable prefix (token) vectors to the keys and values of the self-attention at every layer.
    In detail, two sets of newly initialized prefix vectors $P_i^k, P_i^v \in \mathbb{R}^{l \times d}$ are concatenated with the original key vector and value vector in the self-attention:
    \begin{equation}
    \label{eq:prefix-tuning}
        concat(P_i^k, W_i^k\textbf{h}), concat(P_i^v, W_i^v\textbf{h}).
    \end{equation}
    
    \item Adapter injects two small modules after the self-attention sub-layer and the FFN sub-layer sequentially.
    The adapter module consists of a down-projection, an up-projection and a nonlinear function between them.
    \begin{equation}
    \label{eq:adapter}
        Adapter(\textbf{h}) = \textbf{h} + f(\textbf{h}W_{down})W_{up}, 
    \end{equation} 
    where $\textbf{h}$ is the output from a sub-layer, $W_{down}\in \mathbb{R}^{d\times r}, W_{up}\in \mathbb{R}^{r\times d}$, and $f$ is ReLU.
    
    \item MAM Adapter adds prefix-tuning in the self-attention (i.e., Eq.~\ref{eq:prefix-tuning}) and inserts a parallel adapter module at the FFN side:
    \begin{equation}\label{eq:mam}
        h = Adapter(\textbf{h}) + FFN(\textbf{h})
    \end{equation}
\end{itemize}

\begin{table}[t]
    \centering
    \begin{tabular}{ccc}
    \toprule
    Method & Insertion position & Number of parameters \\
    \midrule
       Bitfit & -  &  $11\times d$   \\
       Prefix-tuning  & attn & $2 \times l \times d$ \\
       Adapter  & attn/ffn & $4 \times r \times d$ \\
       MAM Adapter & attn/ffn & $2\times r \times d + 2 \times l \times d$ \\
       LoRA & attn & $4 \times r \times d$ \\
    \bottomrule
    \end{tabular}
    \caption{Number of parameters used at each layer for different methods. Note that for Bitfit, there are 8 bias terms in each transformer layer and 1 bias term in embedding layer.}
    \label{tab:parameter-count}
\end{table}

\subsubsection{Specification-based}
Specification-based methods only tune certain parameters in the original model.
\begin{itemize}[leftmargin=*]
    \item Bitfit~\cite{zaken2021bitfit} is a very simple method that only trains the bias vectors of the original \acp{PTM} and keeps the rest frozen.
\end{itemize}

\begin{table*}[t]
  \caption{Comparison between full fine-tuning and various parameter-efficient tuning methods using bi-encoder architecture at the retrieval stage.  Best results are marked bold. Note that adding 6.7\% params ($l=400$) for prefix-tuning increases excessive computational cost to document-based tasks which is unacceptable, we thus only experiment with adding 3.6\% params.
  }
        \setlength\tabcolsep{6pt}
    \renewcommand{\arraystretch}{0.6}
  \label{tab:overall retrieval}
  \begin{tabular}{llcccccccc}
  \toprule
    \toprule
    \multirow{2}{*}{Method} & \multirow{2}{*}{\#Params} &  \multicolumn{2}{c}{MARCO Passage} & \multicolumn{2}{c}{TREC2019 Passage} & \multicolumn{2}{c}{MARCO Doc} & \multicolumn{2}{c}{TREC2019 Doc}\\ 
    \cmidrule(lr){3-4} \cmidrule(lr){5-6} \cmidrule(lr){7-8} \cmidrule(lr){9-10} 
     &  & MRR@10 & R@1000 & nDCG@10 & R@100 & MRR@100  & R@100 & nDCG@10  & R@100  \\ 
    \midrule
    Full fine-tuning & 100\% & \textbf{0.316} & \textbf{0.949} & \textbf{0.600} & \textbf{0.715} & \textbf{0.312} & \textbf{0.801} & \textbf{0.462} & \textbf{0.409} \\
    \midrule
    Bitfit & 0.09\% & 0.262 & 0.921 & 0.562 & 0.677 & 0.264 & 0.785 & 0.437 & 0.345\\
    Prefix-tuning & 0.5\% (l=32)& 0.294 & 0.939 & 0.596 &0.692 & 0.266 & 0.782 & 0.423 & 0.326\\
    Adapter & 0.5\% (r=16) & 0.304 & 0.941 & \textbf{0.606} & 0.696 & 0.255 & 0.770 & 0.418 & 0.370\\
    MAM Adapter & 0.5\% (r=16,l=16) & 0.304 & 0.944 & \textbf{0.609} & 0.712 & 0.280 & 0.799 & 0.458 & 0.381 \\
    LoRA & 0.5\% (r=16) &  0.302 & 0.943 & \textbf{0.608} & 0.707 & 0.271 & 0.794 & 0.417 & 0.376\\
    \midrule
    Prefix-tuning & 3.6\% (l=200) & 0.304 & 0.943 & 0.580 & 0.702 & 0.265 & 0.775 & 0.395 & 0.376 \\
    Adapter & 6.7\% (r=200) & 0.316 & 0.946 & 0.587 & 0.687 & 0.270 & 0.785 & 0.433 & 0.400\\
    MAM Adapter  & 6.7\% (r=200,l=200) & 0.314 & 0.947 & \textbf{0.616} & \textbf{0.720} & 0.283 & 0.792 & 0.438 & 0.402\\
    LoRA & 6.7\% (r=200) & 0.316 & 0.946 & 0.597 & 0.715 & 0.279 & 0.794 & 0.417 & 0.379 \\
    \bottomrule
    \bottomrule
  \end{tabular}
\end{table*}

\begin{table*}[t]
    \setlength\tabcolsep{3.8pt}
    \renewcommand{\arraystretch}{0.6}
  \caption{Comparison between full fine-tuning and various parameter-efficient tuning methods using cross-encoder architecture at the re-ranking stage.  Best results are marked bold.
  }
  \label{tab:overall reranking}
  \begin{tabular}{llcccccccc}
  \toprule
    \toprule
    \multirow{2}{*}{Method} & \multirow{2}{*}{\#Params} &  \multicolumn{2}{c}{MARCO Passage} & \multicolumn{2}{c}{TREC2019 Passage} & \multicolumn{2}{c}{MARCO Doc} & \multicolumn{2}{c}{TREC2019 Doc}\\ 
    \cmidrule(lr){3-4} \cmidrule(lr){5-6} \cmidrule(lr){7-8} \cmidrule(lr){9-10} 
     &  & MRR@10 & MRR@100 & nDCG@10 & nDCG100 & MRR@10  & MRR@100 & nDCG@10  & nDCG@100  \\ 
    \midrule
    Full fine-tuning & 100\% & \textbf{0.376} & \textbf{0.383} & \textbf{0.738} & \textbf{0.637} & \textbf{0.404} & \textbf{0.408} & \textbf{0.657} & \textbf{0.536} \\
    \midrule
    Bitfit & 0.09\% & 0.325 & 0.334 & 0.562 & 0.483 & 0.364 & 0.357 & 0.630 & 0.531 \\
    Prefix-tuning & 0.5\% (l=32)& 0.355 & 0.363  &  0.705 & 0.626 & 0.387 & 0.381 & 0.640 & 0.530 \\
    Adapter & 0.5\% (r=16) & 0.366 & 0.371 & 0.714 & 0.626 & 0.397 & 0.392 & 0.653 & 0.534 \\
    MAM Adapter & 0.5\% (r=16,l=16) & 0.365 & 0.373 & 0.717 & 0.629 & 0.390 & 0.395 & 0.632 & 0.531 \\
    LoRA & 0.5\% (r=16) & 0.363 & 0.372 & 0.720 & 0.635 & 0.386 & 0.392 & 0.637 & 0.529 \\
    \midrule
    Prefix-tuning & 3.6\% (l=200)& 0.363 & 0.371 & 0.722 & 0.632 & 0.384 & 0.389 & 0.640 & 0.532\\
    Adapter & 6.7\% (r=200) & 0.373 & 0.381 & 0.735 & 0.637 & 0.402 & 0.407 & 0.631 & 0.528\\
    MAM Adapter  & 6.7\% (r=200,l=200) & 0.369 & 0.380 & 0.731 & 0.633 & 0.397 & 0.402 & 0.630 & 0.528 \\
    LoRA & 6.7\% (r=200) & 0.370 & 0.378 & 0.730 & 0.631 & 0.401 & 0.396 & 0.647 & 0.530\\
    \bottomrule
    \bottomrule
  \end{tabular}
\end{table*}

\subsubsection{Low-rank adaptation}
This type of method hypothesizes that the change of weights during model optimizing has a low intrinsic rank.
Thus, learning a low-rank decomposition matrix for a frozen pre-trained weight matrix can approximate its weight updates, i.e., a fine-tuned pre-trained weight matrix.

\begin{itemize}[leftmargin=*]
    \item LoRA trains rank decomposition matrices, which is a down-project and a up-projection, for the dense layer to approximate the weight updates.
    Specifically, LoRA adds the low-rank matrices to the query and value projection matrices ($W^q, W^v$) in the self-attention.
    Taking $W^q$ as an exmaple:
    \begin{equation}\label{eq:lora}
        \textbf{h} = \textbf{h}\cdot W^q + \Delta W  = \textbf{h}\cdot W^q + s \cdot \textbf{h}\cdot W_{down}W_{up},
    \end{equation}
    where  $s$ is a tunable scalar hyperparameter, $W_{down}\in \mathbb{R}^{d\times r}$, and $W_{up}\in \mathbb{R}^{r\times d}$.
\end{itemize}

We also present the number of parameters used by these methods in Table~\ref{tab:parameter-count}.
Based on this, we can change the number of tunable prefixes $l$ or the hidden size $r$ of the inserted module to control the total number of tunable parameters for fair comparisons.

\section{A Comprehensive Study}
In this section, we conduct a comprehensive study of the above introduced parameter-efficient tuning methods at both the retrieval and re-ranking stages. 
We first analyze the overall experimental performance of existing methods. 
Then, we provide some empirical observations and a following theoretical analysis.


\subsection{Overall Performance}

We conduct experiments on four large-scale standard benchmarks, including MS MARCO passage ranking datasets (MARCO Dev Passage)~\cite{msmarco}, MS MARCO document ranking datasets (MARCO Dev Doc)~\cite{msmarco}, TREC 2019 Deep Learning Track passage ranking task (TREC2019 Passage)~\cite{trec2019}, and TREC 2019 Deep Learning Track document ranking task (TREC2019 Doc)~\cite{trec2019}. 
The detailed experimental setting can be found in Section~\ref{sec:exp_setting}. 
Table~\ref{tab:overall retrieval} and Table~\ref{tab:overall reranking} show the results at the retrieval stage and the re-ranking stage, respectively.

For the retrieval stage, we have the following observations:
(1) Unlike the promising results in NLP, all representative methods cannot achieve a comparable performance over full fine-tuning with less than 1\% of the model parameters on all datasets.
Note that our full fine-tuning baseline is strong as we use mulitple negatives for each query in a mini-batch.
(2) With tuning 6\% of the original model parameters, these methods achieve comparable performance to the full fine-tuning baseline, but still underperform these baselines on the MARCO Passage and TREC2019 Passage.
On MARCO Doc and TREC2019 Doc, they are still very lower than full fine-tuning since we train the dense retrieval models with BM25 negatives which is a little weaker.
Existing works like ANCE~\cite{ance} and ADORE~\cite{Zhan2021adore} always use the checkpoint trained on the MARCO Passage as the starting point for MARCO Doc.
But this makes for an unfair comparison since the parameter-efficient tuning methods would have different starting points.
We will do further comparisons by training the dense retrieval models with hard negatives in Section~\ref{sec:exp_hn}.
(3) Among these parameter-efficient tuning methods, Bitfit performs worst, while LoRA, Adapter and MAM Adapter are more effective than prefix-tuning. 
Prefix-tuning increases computational cost as it prepends additional trainable tokens in the hidden layer.

For the re-ranking stage, we can see that:
(1) The relative order of different parameter-efficient tuning methods at this stage is quite consistent with that at the retrieval stage. 
(2) Our finding is not consistent with \citet{Jung2022sime-siamese} in which they found that prefix-tuning and LoRA are able to outperform the full fine-tuning on small datasets including Robust04 and ClueWeb09, and nonstandard MARCO document ranking dataset with less than 1\% of model parameters.
In our experiments, with more strong baselines (i.e., training cross-encoder with several negatives in a mini-batch), these parameter-efficient tuning methods cannot outperform the full fine-tuning on standard large-scale datasets.

\begin{figure}[t]
    \centering
    \includegraphics[scale=0.32]{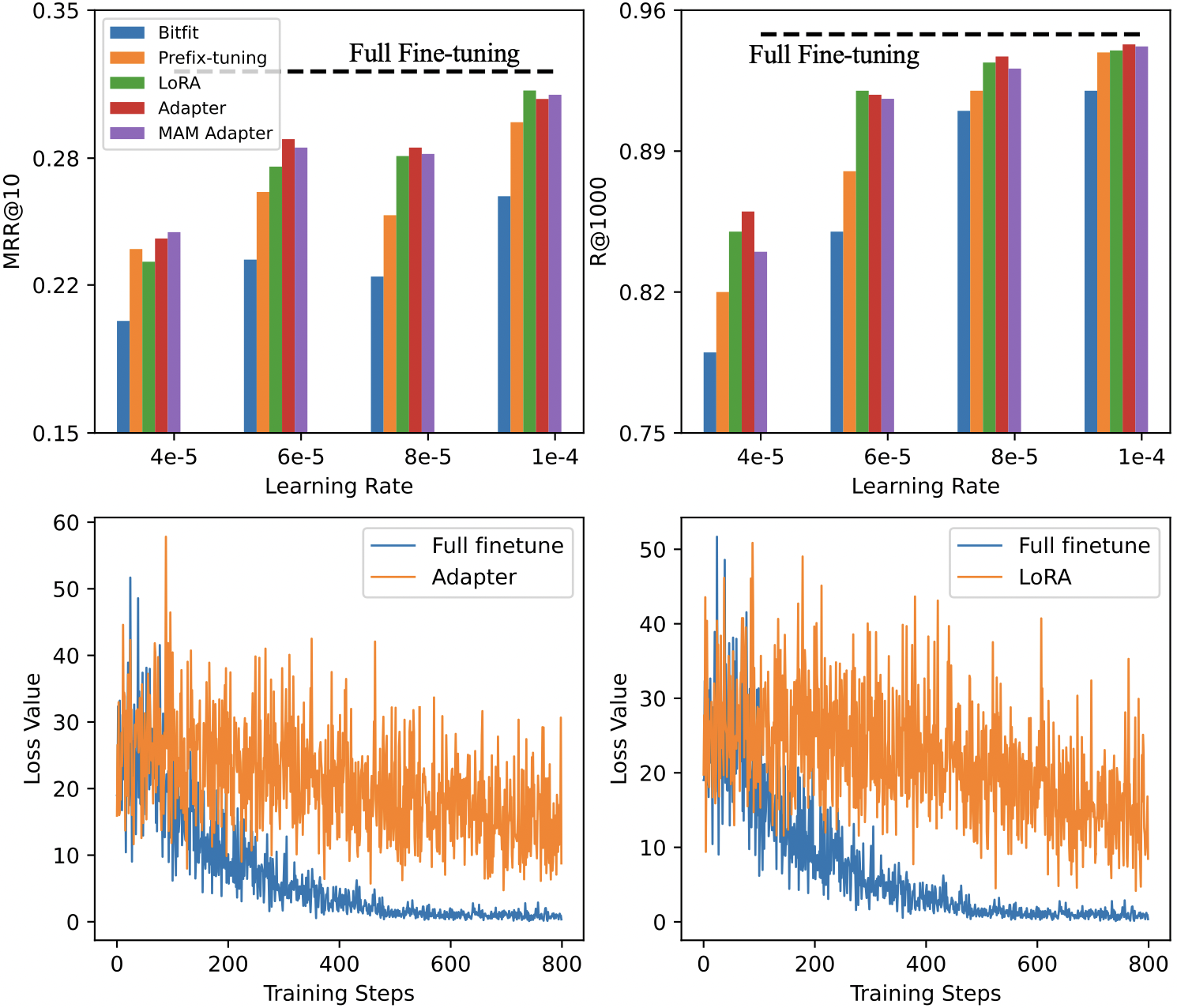}
    \caption{Top: The retrieval performance of various parameter-efficient tuning methods using different learning rates on MARCO Passage. Bottom: The loss value of full fine-tuning and the two best performing parameter-efficient methods (i.e., Adapter and LoRA) over training steps.}
    \label{fig:observation}
\end{figure}

\subsection{Empirical Observation}

Besides the above overall performance, we provide some empirical observations about the training and convergence of existing methods. 
We find that these methods are very sensitive to hyperparameters, such as learning rate.
We also noticed that the loss value at the early training stage is very high which seems to be hard to converge.
We take the MARCO Passage as an example and other datasets have the same observation.
We show the results at the retrieval stage with different learning rates, i.e., ranging from 4e-5 to 1e-4.
The detailed experimental setting can be found in Section~\ref{sec:exp_setting}.

As shown in Figure~\ref{fig:observation}, the performance of these parameter-efficient tuning methods varies wildly with different learning rates.
A low learning rate always performs worse than a high learning rate in terms of both MRR@10 and R@100 metrics.
The results implicate that low learning rates may not find a good optimization direction and are difficult to skip the local optima, leading to slower convergence. 
Then, we take a look at the loss curve of these methods in the early training stage.
As shown at the bottom of Figure~\ref{fig:observation}, we can see that the two best performing parameter-efficient methods Adapter and LoRA have a higher loss value compared to full fine-tuning and their loss values fluctuate wildly, ranging from about 50 to 10. 
The unstable training process and slow convergence indicate that these methods may be essentially hard to optimize.


\subsection{Theoretical Analysis}

We theoretically show why the standard setting of existing parameter-efficient tuning methods is not learning-efficient. 

For Transformer-based \acp{PTM}, each layer contains a multi-head attention layer (MH), a FFN layer and two RCLN functions.
As we introduced in Section~\ref{sec:pet}, most of the parameter-efficient tuning methods inject modules to the FFN sub-layer and MH sub-layer in an inside manner, e.g., an Adapter following the MH and another following the FFN. 

For simplicity, this can be treated as some form of modifications to FFN and MH, i.e., $f(MH(\cdot)), g(FFN(\cdot))$.
Thus, these parameter-tuning methods can be formulated as follows:
\begin{equation}\label{eq:pet-transformer}
     h=RCLN(g(FFN(RCLN(f(MH(x)))))),
\end{equation}
where $f, g$ are the inserted module, such as the Adapter module in Eq.~\ref{eq:adapter}, the LoRA module in Eq.~\ref{eq:lora}, the prefix module in Eq.~\ref{eq:prefix-tuning} and the MAM Adapter module in Eq.~\ref{eq:mam}.

During training, the goal is to minimize the loss over every training example $x=(q, d)$. 
The model parameters $\Theta$ of step $t$ are optimized by gradient descent methods (GD):
\begin{equation}
    \Theta_{t+1} = \Theta_{t} - \eta \nabla(J(x, y ; \Theta_{t}),
\end{equation}
where $\eta$ is the learning rate, $y$ is the label, and $J$ is the loss function.
For simplicity, we will omit the complex loss functions used in the ranking task here.
According to the chain rule, the gradient on step $t$ in Eq.~\ref{eq:pet-transformer} is computed as follows:
\begin{equation}
    \frac{dJ}{dx} = \frac{dJ}{dRCLN_t} \frac{dRCLN_t}{dg_t}\frac{dg_t}{dFFN_t}\frac{dFFN_t}{dRCLN_t}\frac{dRCLN_t}{df_t}\frac{df_t}{dMH_t}\frac{dMH_t}{dx_t}.
\end{equation}
Since the parameters of $RCLN_t$, $FFN_t$ and $MH_t$ are kept frozen, only $f_t, g_t$ are updated to:
\begin{equation}\label{eq:fg}
\begin{split}
    g_{t+1} =& g_t - \eta \nabla(g) = g_t - \eta \frac{dJ(x,y; \Theta)}{dg}
    = g_t - \eta \frac{dJ}{dRCLN_t} \frac{dRCLN_t}{dg_t}, \\ 
    f_{t+1} =& f_t - \eta \nabla(f) = f_t - \eta \frac{dJ(x,y; \Theta)}{df} \\
    =& g_t - \eta \frac{dJ}{dRCLN_t} \frac{dRCLN_t}{dg_t}\frac{dg_t}{dFFN_t}\frac{dFFN_t}{dRCLN_t}\frac{dRCLN_t}{df_t}, 
\end{split}
\end{equation}

As we can see, the gradients of $f, g$ are computed based on the frozen parameters including $RCLN_t$, $FFN_t$ and $MH_t$.
The ideal gradient descent direction is $\nabla(\Theta_{t})$ including all parameters, but the actual gradient update direction is only  $\nabla(f_t,g_t))$.
So there remains a discrepancy between the ideal optimization direction and the actual update direction
\begin{equation*}
    \delta = \nabla(\Theta_{t}) - \nabla(f_t,g_t)).
\end{equation*}
The MH (or FFN) may be the main contribution module to the gradient of the $\nabla(\Theta_{t})$, that is, updating the parameters of MH (or FFN) may greatly decrease the loss value of the input batch.
But only $f_t,g_t$ are updated, so this can explain why the loss value of parameter-efficient tuning methods varies wildly during training, since $f_t,g_t$ may contribute little to the gradient of the $\nabla(\Theta_{t})$.
Therefore, the separation of these inserted trainable parameters leads to the discrepancy problem which can make the optimization difficult and hurt the performance.


\begin{figure*}
    \centering
    \includegraphics[scale=0.38]{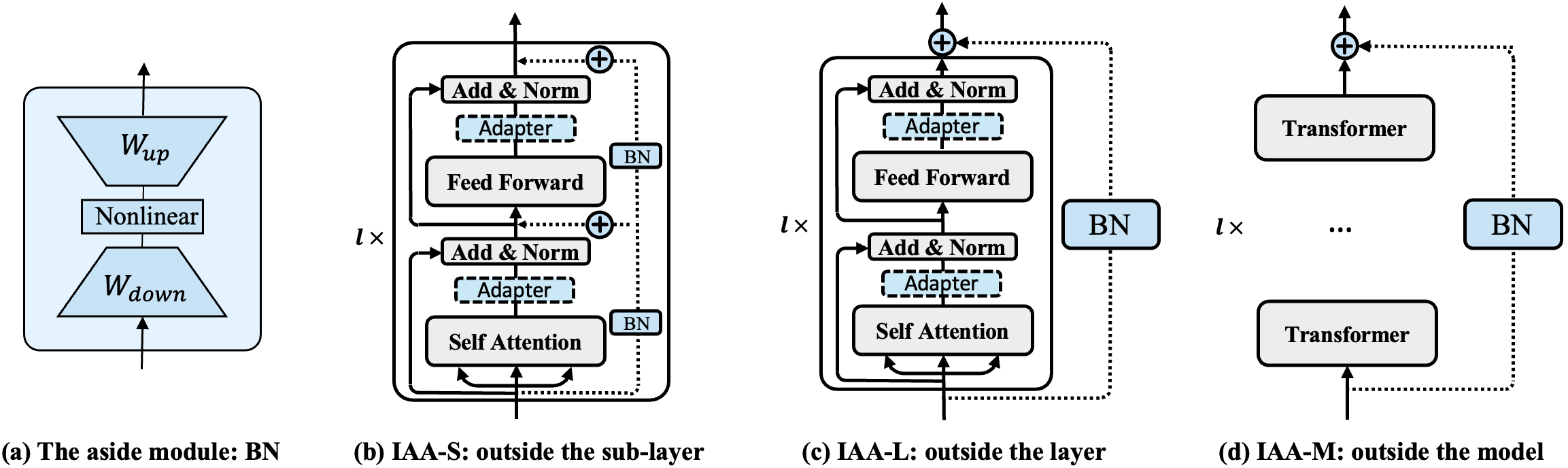}
    \caption{The aside module and three variants with different  insertion ways. In this way, all the extra inserted modules can form a pathway. $l$ denotes the number of Transformer.}
    \label{fig:ours}
\end{figure*}

%% file: method.tex
\section{Our Method}

Our analysis shows that these scattered modules in an inside manner lead to unsmooth transferring of updatable gradients. Inspired by skip connection~\cite{he2016resnet}, beyond the inside manner, we propose to inject additional modules alongside the PTMs to create a pathway for updatable gradients.
Specifically, in this way, the scattered modules can be directly connected throughout the whole \acp{PTM}.
Formally, we denote this type of module as the \textit{aside} module and the module which is injected into the model as the \textit{inside} module.
Without the effect of the original frozen model parameters, these aside modules can create an unimpeded path to make the updatable gradients flow fluently.

The aside module is denoted as $z(x)$.
So, according to Eq.~(\ref{eq:fg}), $ \nabla(z)$ is better than $\nabla(f)$ and $\nabla(g)$ since its gradient is only based on the final loss $J$ and itself, i.e., $\nabla(z)=\frac{dJ}{dz}$, and don't have to multiply the gradients of RCLN, FFN and MH.
In this way, the aside module is updated without the barrier of frozen model parameters and thus mitigates the optimization discrepancy.

Although the inside module suffers from the optimization discrepancy, it's more expressive and has a bigger impact on the final output than the aside module since its output will be transformed by the next following complex modules like MH and FFN.
To leverage the merits of these two kinds of modules, we propose to combine the inside module and the aside module for better performance.

\textbf{The Inside Module} 
We can adopt any parameter-efficient tuning methods which inject new parameters into \acp{PTM}, as our inside module.
In our pilot experiments, we find that Adapter-based and LoRA perform best across all parameter-efficient tuning methods.
So in our main experiments, we employ Adapter as our inside module, and we also conduct experiments with LoRA in Section~\ref{sec:compare_ss}.
We leave the study of choosing or designing the inside module for future work.

\textbf{The Aside Module}
As shown in Figure~\ref{fig:ours}, our proposed aside module is a bottleneck architecture (BN) containing a down-projection, a nonlinear function and an up-projection.
Compared to Adapter, BN has no residual connection, and compared to LoRA, BN adds a nonlinear function.
As depicted in Figure~\ref{fig:ours}, we investigate three ways of inserting BN to the model along with the inside modules, i.e., outside the sub-layer, outside the layer, and outside the model.
We denote these three Inside and Aside (IAA) structures as IAA-S, IAA-L, and IAA-M respectively.
\begin{itemize}[leftmargin=*]
    \item \textbf{IAA-S} inserts two BN modules outside the two sub-layer in Transformer, i.e., FFN sub-layer and MH sub-layer.
    Note that the output of each BN is added to the output of the residual connection and layer normalization.
    The number of parameters is computed as $4\times r \times d \times l$.
    \item \textbf{IAA-L} inserts one BN module outside the Transformer layer and there are $l$ modules for a whole \acp{PTM}.
    The number of parameters is computed as $2\times r \times d \times l$.
    \item \textbf{IAA-M} inserts one BN modules outside the whole \acp{PTM}.
    It takes the output of the embedding layer and adds its output to the final output of the whole model.
    The number of parameters is computed as $2\times r \times d$.
\end{itemize}
In order to fairly compare these structures, we can control the hidden size $r$ to keep the number of parameters in each structure the same.
As the BN gets farther from the original Transformer, it can have a larger hidden size to have the same number of parameters.

%% file: exp-settings.tex
\begin{table*}[t]
\renewcommand{\arraystretch}{0.6}
  \caption{Comparisons between IAA and the baselines at the retrieval stage. Two-tailed t-tests demonstrate the improvements of IAA over baselines are statistically significant ( $p \le 0.05$).
  $\ast$ indicate significant improvements over full fine-tuning. $\dag$ indicate significant improvements over best parameter-efficient tuning methods (PET) at the same setting.}
  \label{tab:dense-retrieval}
  \begin{tabular}{llllllllll}
  \toprule
    \toprule
    \multirow{2}{*}{Method} & \multirow{2}{*}{\#Params} &  \multicolumn{2}{c}{MARCO Passage} & \multicolumn{2}{c}{TREC2019 Passage} & \multicolumn{2}{c}{MARCO Doc} & \multicolumn{2}{c}{TREC2019 Doc}\\ 
    \cmidrule(lr){3-4} \cmidrule(lr){5-6} \cmidrule(lr){7-8} \cmidrule(lr){9-10} 
     &  & MRR@10 & R@1000 & nDCG@10 & R@100 & MRR@100  & R@100 & nDCG@10  & R@100  \\ 
    \midrule
    Full fine-tuning & 100\% & 0.316 & 0.949 & 0.600 & 0.715 & \textbf{0.312} & \textbf{0.801} & \textbf{0.462} & \textbf{0.409} \\
    \midrule
    Best PET & 0.5\% & 0.304 & 0.944 &  0.609 & 0.712 & 0.280 & 0.799 & 0.458 & 0.381 \\
    IAA-S Adapter & 0.5\% (r=8,ar=8) & 0.312$^\dag$ & 0.941 & 0.605 & 0.719 & 0.285 & 0.785 & 0.454 & 0.384 \\
    IAA-L Adapter & 0.5\% (r=12,ar=12) & 0.314$^\dag$ & 0.943 & 0.615$^\dag$ & 0.735$^\ast$ & 0.292 & 0.792 & 0.446 & 0.391 \\
    IAA-M Adapter & 0.5\% (r=15,ar=24) & 0.309 & 0.941 & 0.602 & 0.721 & 0.287 & 0.782 & 0.449 & 0.385\\
    \midrule
    Best PET & 6.7\% & 0.316 & 0.946 & 0.616 & 0.720 & 0.283 & 0.792 & 0.438 & 0.402\\
    IAA-S Adapter & 6.7\% (r=100,ar=100) & 0.324 & 0.947 & 0.581 &	0.719 & 0.290 & 0.798 & 0.441 & 0.398\\
    IAA-L Adapter & 6.7\% (r=50,ar=300) & \textbf{0.327$^{\dag\ast}$} & \textbf{0.951} & \textbf{0.617$^{\ast}$} & \textbf{0.735$^\dag$} &  0.295$^\dag$ & 0.795 & 0.439 & 0.395 \\
    IAA-M Adapter & 6.7\% (r=185,ar=960) & 0.321 & 0.948 & 0.592 & 0.710 & 0.285 & 0.793 & 0.437 & 0.402\\
    \bottomrule
    \bottomrule
  \end{tabular}
\end{table*}

\section{EXPERIMENTAL SETTINGS}\label{sec:exp_setting}

In this section, we introduce our experimental settings, including datasets,  baseline methods, evaluation metrics, and training details.

\subsection{Datasets}
We conduct our experiments on 4 standard ranking datasets, including MS MARCO passage ranking datasets (MARCO Passage)~\cite{msmarco}, MS MARCO document ranking datasets (MARCO Doc)~\cite{msmarco}, TREC 2019 Deep Learning Track passage ranking task (TREC2019 Passage)~\cite{trec2019}, and TREC 2019 Deep Learning Track document ranking task (TREC2019 Doc)~\cite{trec2019}.
MARCO Passage contains 0.5 million training queries, 6 thousand dev queries and 8.8 million passages.
MARCO Doc contains 0.4 million training queries, 5 thousand dev queries and 3 million documents.
For these two MARCO datasets, we report the performance on dev set following existing work~\cite{ance,Zhan2021adore,nogueira2019bert-pas-ranking,Ma2021prop}.
The two TREC2019 datasets share the same training set and document collection with their corresponding MARCO datasets, but they have a fine-grained test set containing 200 queries.

\subsection{Baselines}

We use the BERT-base model as the backbone, and use the cross-encoder architecture and the bi-encoder architecture for re-ranking and dense retrieval, respectively.

Our baseline includes the full fine-tuning and 5 representative parameter-efficient tuning methods as we introduced in Section~\ref{sec:pet}, including BitFit~\cite{zaken2021bitfit}, prefix-tuning~\cite{prefix-tuning}, Adapter~\cite{houlsby2019adapter}, MAM Adapter~\cite{he2021mam-adapter} and LoRA~\cite{lora}.
The recent proposed Semi-Siamese method~\cite{Jung2022sime-siamese} is applied to the prefix-tuning and LoRA in their experiment.
The Semi-Siamese prefix-tuning (SS prefix), besides the common prefix, uses some specific prefixes for the query and the document respectively to model their distinct characteristics.
The Semi-Siamese LoRA (SS LoRA) use the same query weight matrices and different value weight matrices for the query and the document

\subsection{Evaluation Metrics}
We report the official metrics of these four benchmarks. 
For the MARCO Passage, we report the Mean Reciprocal Rank at 10 (MRR@10) and recall at 1000 (R@1000). 
For the MARCO Doc, we report the MRR@100 and R@100. 
For TREC2019 Passage, we report normalized discounted cumulative gain at 10 (NDCG@10), and R@1000 while for TREC2019 Doc, we report NDCG@10 and R@100.

\subsection{Training and Optimization}

For the cross-encoder model which is used for the re-ranking stage, the query and the document are concatenated into a single sequence to input to the model.
We truncate the sequence to the first 128 tokens and 512 tokens for passage datasets and document datasets, respectively.
We use cross-entropy pairwise loss and pair 5 negative examples for each query in a mini-batch.
We use the official top-k candidates as the negatives.
We use a batch size of 72 and 36 for passage datasets and document datasets, respectively.
We train 5 epochs for all methods and choose the best checkpoint.
The only difference between full fine-tuning and other baselines is that we set different learning rates.
For full fine-tuning, we use a learning rate of 2e-5.
For all other parameter-efficient tuning methods, we use a learning rate of 1e-4.

For the bi-encoder model which is used for dense retrieval, the query and the document are encoded separately.
We set the maximum length of the query to 32, the passage to 128, and the document to 512.
We use the official top-k candidates for the passage retrieval task and use BM25 top-k candidates retrieved by anserini~\cite{Yang2017anserini} for document retrieval task~\cite{Yang2017anserini}.
Training dense retrieval models with official top-k candidates on MARCO Doc results in bad performance.
We pair 7 negative examples for each query on passage retrieval and 1 negative example on document retrieval.
We use a batch size of 64 and 44 for passage datasets and document datasets, respectively.
We train 3 epochs, and 6 epochs for passage datasets and document datasets, respectively.
For full fine-tuning, we use a learning rate of 2e-5.
For all parameter-efficient tuning methods, we use a learning rate of 1e-4. 
For all experiments, we use the Adam optimizer with a linear warm-up over the first 10\% steps.

%% file: exp-results.tex
\section{Empirical Results}
In this section, we report and analyze the experimental results to demonstrate the effectiveness of the proposed method. We target the following research questions: 
\begin{itemize}[leftmargin=*]
\item \textbf{RQ1:} How does our method perform compared with full fine-tuning and other parameter-efficient tuning methods?
\item \textbf{RQ2:} How does our method perform compared with the Semi-Siamese bi-encoder neural models on the re-ranking stage?
\item \textbf{RQ3:} How does our method perform compared with advanced dense retrieval models when training with hard negatives?
\item \textbf{RQ4:} How does the hidden size of the aside module affect the performance?
\item \textbf{RQ5:} How does the connected modules affect the optimization process?
\end{itemize}

\begin{table*}[t]
\renewcommand{\arraystretch}{0.6}
  \setlength\tabcolsep{2.5pt} 
  \caption{Comparisons between IAA and the baselines on the re-ranking stage. Two-tailed t-tests demonstrate the improvements of IAA over baselines are statistically significant ( $p \le 0.05$).
  $\ast$ indicate significant improvements over full fine-tuning. $\dag$ indicate significant improvements over best parameter-efficient tuning methods (PET) at the same setting.}
  \label{tab:main_reranking}
  \begin{tabular}{llllllllll}
  \toprule
    \toprule
    \multirow{2}{*}{Method} & \multirow{2}{*}{\#Params} &  \multicolumn{2}{c}{MARCO Passage} & \multicolumn{2}{c}{TREC2019 Passage} & \multicolumn{2}{c}{MARCO Doc} & \multicolumn{2}{c}{TREC2019 Doc}\\ 
    \cmidrule(lr){3-4} \cmidrule(lr){5-6} \cmidrule(lr){7-8} \cmidrule(lr){9-10} 
     &  & MRR@10 & MRR@100 & nDCG@10 & nDCG100 & MRR@10  & MRR@100 & nDCG@10  & nDCG@100  \\ 
    \midrule
     Full fine-tuning & 100\% & 0.376 & 0.383 & 0.738 & 0.637 & 0.404 & 0.408 & 0.657 & 0.536 \\
    \midrule
    Best PET & 0.5\% & 0.366 & 0.371 & 0.720 & 0.635 & 0.397 & 0.392 & 0.653 & 0.534 \\
    IAA-S Adapter & 0.5\% (r=8,ar=8) & 0.371 & 0.377 & 0.731$^{\dag}$ &	0.632 & 0.395 & 0.393 & 0.655 & 0.533 \\
    IAA-L Adapter & 0.5\% (r=12,ar=12) & 0.373$^{\dag}$ & 0.379$^{\dag}$ & 0.732$^{\dag}$ & 0.633 & 0.399 & 0.403$^{\dag}$  & 0.656 & 0.537 \\
    IAA-M Adapter & 0.5\% (r=15,ar=24) &  0.369 & 0.373 & 0.725 & 0.630 & 0.393 & 0.391 & 0.652 & 0.531 \\
    \midrule
    Best PET & 6.7\% & 0.373 & 0.381 & 0.735 & 0.637 & 0.402 & 0.407 & 0.647 & 0.530 \\
    IAA-S Adapter & 6.7\% (r=100,ar=100) & 0.382$^{\dag}$ & 0.385 & \textbf{0.742} &	0.635 & 0.408 & 0.412 & 0.651 & 0.535 \\
    IAA-L Adapter & 6.7\% (r=50,ar=300) & \textbf{0.385$^{\ast\dag}$} & \textbf{0.392$^{\ast\dag}$} & 0.740 & \textbf{0.639} & \textbf{0.412$^{\dag}$} & \textbf{0.414} & \textbf{0.657$^{\dag}$} & \textbf{0.538}\\
    IAA-M Adapter & 6.7\% (r=185,ar=960) & 0.379 & 0.384 & 0.739 & 0.636 & 0.404 & 0.410 & 0.649 & 0.529\\
    \bottomrule
    \bottomrule
  \end{tabular}
\end{table*}

\subsection{Main Results}

To answer \textbf{RQ1}, we compare three variants of IAA Adapter with full fine-tuning and the best parameter-efficient tuning methods on four standard large-scale datasets.
Table~\ref{tab:dense-retrieval} and Table~\ref{tab:main_reranking} show the results at the retrieval stage and the re-ranking stage, respectively.

We first look at the results at the retrieval stage:
(1) Our best IAA model with tuning less than 1\% of the model parameters achieve a comparable performance over full fine-tuning, and is significantly better than the best PET on some datasets like MARCO Passage.
(2) By tuning 6.7\% of the model parameters, our best model could outperform the full fine-tuning baseline on two passage retrieval datasets.
On MARCO Passage, it's also significantly better than full fine-tuning in terms of MRR@10.
This demonstrates that by introducing the connected aside module, our method is able to improve the performance.
On two document retrieval tasks, we find that our methods cannot outperform the full fine-tuning baseline indicting training with BM25 negatives is not enough for bi-encoder on document retrieval.
We leave this for further study.
(3) Compare the three insertion structures, we find that IAA-L which injects the aside module outside the layer performs best.
One possible reason is that IAA-S which injects the aside module outside the sub-layer has a smaller hidden size of the inside module than IAA-L which may limit its capacity.
For IAA-M, although it have bigger hidden size for the aside module, its representative power is not as good as IAA-L since the output of each aside module in IAA-L can be transformed by the original parameters.

We then look at the re-ranking stage, and we find that the performance trend on the re-ranking stage is consistent with the retrieval stage:
(1) Our method is significantly better than the best parameter-efficient tuning methods in terms of MRR@10 on MARCO Passage and MARCO Doc.
(2) All types of IAA can outperform the full fine-tuning baseline by tuning 6.7\% of the model parameters, indicating the effectiveness of IAA.
(3) Unlike the poor performance on document retrieval tasks, IAA could outperform the full fine-tuning with 6.7\% of the model parameters.
It demonstrates that applying parameter-efficient tuning methods on cross-encoder perform better than on the bi-encoder.

\subsection{Comparison with Semi-Siamese Bi-encoder Baseline}\label{sec:compare_ss}

To answer \textbf{RQ2}, we compare our method with the recently proposed Semi-Siamese methods, i.e., SS prefix-tuning and SS LoRA~\cite{Jung2022sime-siamese}.
These two methods can only apply to the bi-encoder architecture and they leverage this method at the re-ranking stage.
As our method is a general method, thus we utilize a bi-encoder architecture on the re-ranking stage for a fair comparison.
We also use IAA-S LoRA which uses LoRA as the inside module to compare with SS LoRA.
Experiments are conducted on MARCO Passage with only tuning 0.5\% of the model parameters.
The results are shown in the table~\ref{tab:ss-comp}.
We can see the SS prefix-tuning performs worst and this is consistent with our previous findings where prefix-tuning is not as  effective as LoRA and Adapter-based.
Our methods including IAA-L LoRA and IAA-L Adapter, are significantly better than the two baseline methods indicating the aside module is more useful and effective for model training.

\begin{table}[t]
    \centering
    \begin{tabular}{llllll}
    \toprule
    \toprule
       \multirow{2}{*}{Model} &  \multicolumn{1}{c}{MARCO Passage} & \multicolumn{1}{c}{MARCO Doc}  \\
      \cmidrule(lr){2-2} \cmidrule(lr){3-3} 
   & MRR@10  & MRR@100  \\
    \midrule
       SS prefix-tuning & 0.342 & 0.375
       \\
       SS LoRA & 0.351 & 0.383 \\
       IAA-L LoRA & 0.366$^{\dag\ast}$ & \textbf{0.391$^{\dag\ast}$} \\
       IAA-L Adapter & \textbf{0.367}$^{\dag\ast}$ & 0.389$^{\ast}$ \\
    \bottomrule
    \bottomrule
    \end{tabular}
    \caption{Performance comparison with Semi-Siamese using a bi-encoder architecture on the re-ranking stage. 
    Two-tailed t-tests demonstrate the improvements are statistically significant ( $\ast,\dag$ indicates $p \le 0.05$ over SS prefix-tuning and SS LoRA, respectively).}
    \label{tab:ss-comp}
\end{table}

\subsection{Comparison with Advanced Dense Retrieval Models by Training with Hard Negatives}\label{sec:exp_hn}

To answer \textbf{RQ3}, we train the dense retrieval models using hard negatives for the parameter-efficient tuning methods.
Following STAR~\cite{Zhan2021adore}, we mine the static hard negatives using BM25 warm-up checkpoint and train the dense retrieval model on hard negatives for another 2-3 epochs.
As shown in Table~\ref{tab:dr_hn}, we can observe that by training with hard negatives, parameter-efficient tuning methods achieve comparable performance over some advanced dense retrieval models such as ANCE, and ADORE.
And our proposed IAA-L Adapter can still outperform full fine-tuning baseline and is significantly better than other parameter-efficient tuning methods such as Adapter and LoRA.
We could see that compared with RocketQA which utilizes several training techniques like cross-batch training, denoising false negatives, and data augmentation, all parameter-efficient tuning methods are still far behind it.

\begin{table}[t]
  \caption{Comparison with advanced dense retrieval models by training PET with hard negatives on the MARCO Passage. Best results are marked bold.}
  \label{tab:dr_hn}
  \begin{tabular}{lccc}
    \toprule
  Model &  MRR@10 & R@1000  \\
    \midrule
    ANCE\cite{ance} & 0.330 & 0.959  \\
    TCT-ColBERT\cite{Lin2021tct-colbert} & 0.335 & 0.964 \\
    TAS-B\cite{Hofsttter2021tas} & 0.343 & \textbf{0.976} \\
    ADORE+STAR\cite{Zhan2021adore} & 0.347 & - \\
    RoctetQA \cite{Qu2021rocketqa} &  \textbf{0.367} & - \\
    \midrule
    full fine-tuning & 0.341 & 0.961 \\
    Adapter & 0.334 & 0.953\\
    MAM Adapter & 0.332 & 0.959\\
    LoRA & 0.331 & 0.957 \\
    IAA-L Adapter & 0.343 & 0.971 \\
  \bottomrule
\end{tabular}
\end{table}

\subsection{Impact of Hidden Size of the Aside Module}
To answer \textbf{RQ4}, we conduct an analysis to investigate the impact of the hidden size of the aside model.
We experiment on MARCO Passage under the dense retrieval setting.
We vary the hidden size of the aside model but still keep the total number of tuning parameters fixed.
That is, the larger the aside module, the smaller the inside module and vice versa.

\begin{figure}[t]
    \centering
    \includegraphics[scale=0.4]{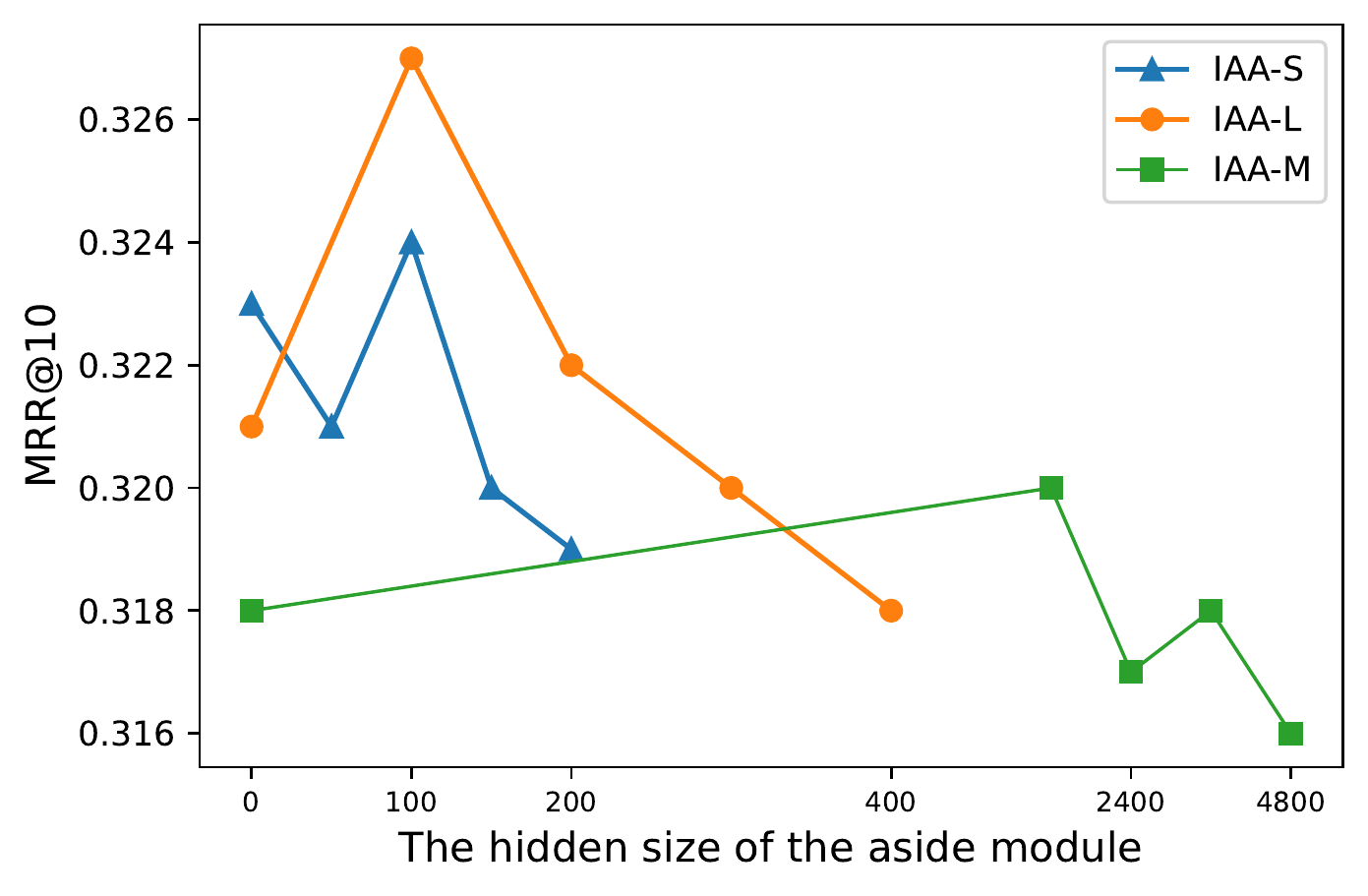}
    \caption{The impact of the hidden size of the aside module.}
    \label{fig:ablation}
\end{figure}

As shown in Figure~\ref{fig:ablation}, we can see that different sizes have a big impact on the performance.
When the hidden size of the aside model is 0, it degrades to the skip connection which only connects the input and output of the original model.
In this setting, IAA-S performs better than IAA-L and IAA-M indicating that a fine-grained skip connection is better than a coarser-grained.
When the hidden size of the inside module is 0, it becomes a totally parallel aside module.
We can see that the aside module underperforms the inside module with skip connection, i.e., the hidden size of the aside module is 0.
This verifies our hypothesis that the inside module is more expressive and has a larger capacity than the aside module.
One possible reason is that the output of the inside module will be transformed by the next following complex Transformer modules like multi-head attention.

\subsection{Convergence Analysis}

\begin{figure}[t]
    \centering
    \includegraphics[scale=0.4]{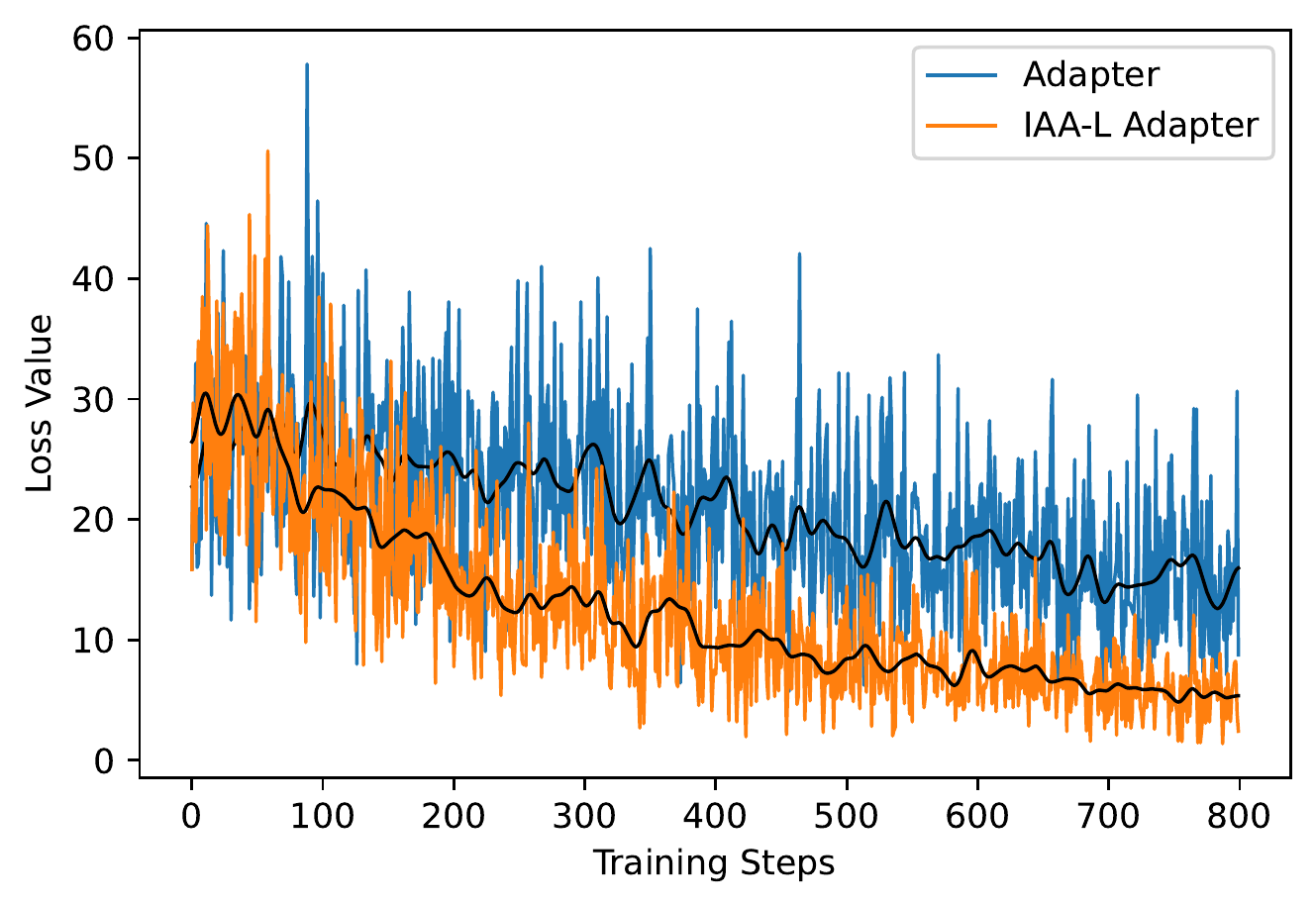}
    \caption{The loss value over training steps.}
    \label{fig:convergence}
\end{figure}

To answer \textbf{RQ5}, we visualize the training loss Adapter and IAA-L Adapter on MARCO Passage at the retrieval stage.
As shown in Figure~\ref{fig:convergence}, IAA-L Adapter has a lower loss value than Adapter and also converges faster than Adapter.
This demonstrates that by adding the aside module, IAA-L Adapter could alleviate the optimization discrepancy problem which is caused by the separation of the trainable modules.
One possible reason is that the aside module eases optimization and accelerates training convergence by smoothing the loss surface.
This has been verified by the ~\cite{li2018visualizing} which says skip connections could promote flat minimizers and prevent the transition to chaotic behavior.


%% file: related.tex
\section{Related Work}

In this section, we briefly review the fine-tuning approaches for \acp{PTM} in IR. 
Fully fine-tuning large \acp{PTM} like BERT~\cite{Devlin2019BERT} is the widely used approach in IR, since it achieve strong performance at both the retrieval stage~\cite{Zhan2021adore,ance,Karpukhin2020dpr,Chen2022GEREGE} and the re-ranking stage~\cite{Ma2021prop,nogueira2019bert-pas-ranking}.
Another approach is the feature-based as used in ELMo~\cite{Peters2018elmo}.
The pre-trained representations input to task-specific architectures as features. 
CEDR~\cite{MacAvaney2019cedr} has investigated this approach in several TREC datasets and found the performance of feature-based degrades greatly compared with the fully fine-tuning.
\citet{Jung2022sime-siamese} firstly apply prefix-tuning and LoRA to the re-ranking stage.
They found that the two kinds of parameter-efficient methods can outperform the full fine-tuning on small test data.
But with more strong baselines, our findings are not consistent with theirs and we propose a more universal method which can be applied to various architectures and parameter-efficient tuning methods.